\documentclass{egpubl_pg_WIP}
\usepackage{pg2021w}
 
\WsPaper           % uncomment for final version of EG Workshop contribution
\usepackage[T1]{fontenc}
\usepackage{dfadobe}  

\usepackage{cite}  % comment out for biblatex with backend=biber
\BibtexOrBiblatex
\electronicVersion
\PrintedOrElectronic
\ifpdf \usepackage[pdftex]{graphicx} \pdfcompresslevel=9
\else \usepackage[dvips]{graphicx} \fi

\usepackage{egweblnk} 
\usepackage{amsmath}

\usepackage{algorithm2e}
\SetKwFunction{saturate}{saturate}

\title{Cloud-Assisted Hybrid Rendering for Thin-Client Games and VR Applications}

\author[Tan et al.]
{\parbox{\textwidth}{\centering 
Tan Yu Wei\orcid{0000-0002-7972-2828}, 
Louiz Kim-Chan\orcid{0000-0001-8981-0592}, 
Anthony Halim\orcid{0000-0002-4340-5820}, 
Anand Bhojan\orcid{0000-0001-8105-1739}
}
\\
{\parbox{\textwidth}{\centering School of Computing, National University of Singapore}}}
\begin{document}

% uncomment for using teaser
% \teaser{
%  \includegraphics[width=\linewidth]{figures/Teaser}
%  \centering
%   \caption{We perform ray tracing on the server to generate a visibility bitmap for three scene lights (left). For each pixel, its material colour is shown if the light is not obstructed from its world space position. The bitmap which contains ray-traced shadow information is combined with Lambertian shading at thc client to form the final output (right). Our test scene is \href{https://www.blendswap.com/blend/13491}{\textsc{The Modern Living Room}} (\href{https://creativecommons.org/licenses/by/3.0/}{CC BY}).}
% \label{fig:teaser}
% }

\maketitle
%-------------------------------------------------------------------------
\begin{abstract}

We introduce a novel distributed rendering approach to generate high-quality graphics in thin-client games and VR applications. Many mobile devices have limited computational power to achieve ray tracing in real-time. Hence, hardware-accelerated cloud servers can perform ray tracing instead and have their output streamed to clients in remote rendering. Applying the approach of distributed hybrid rendering, we leverage the computational capabilities of both the thin client and powerful server by performing rasterization locally while offloading ray tracing to the server. With advancements in 5G technology, the server and client can communicate effectively over the network and work together to produce a high-quality output while maintaining interactive frame rates. Our approach can achieve better visuals as compared to local rendering but faster performance as compared to remote rendering.

\begin{CCSXML}
	<ccs2012>
		<concept>
			<concept_id>10010147.10010371.10010372</concept_id>
			<concept_desc>Computing methodologies~Rendering</concept_desc>
			<concept_significance>500</concept_significance>
		</concept>
		<concept>
			<concept_id>10010147.10010371.10010372.10010374</concept_id>
			<concept_desc>Computing methodologies~Ray tracing</concept_desc>
			<concept_significance>500</concept_significance>
		</concept>
		<concept>
			<concept_id>10010405.10010476.10011187.10011190</concept_id>
			<concept_desc>Applied computing~Computer games</concept_desc>
			<concept_significance>500</concept_significance>
		</concept>
	</ccs2012>
\end{CCSXML}

\ccsdesc[500]{Computing methodologies~Rendering}
\ccsdesc[500]{Computing methodologies~Ray tracing}
\ccsdesc[500]{Applied computing~Computer games}

\printccsdesc   
\end{abstract}

\maketitle
 
\section{Introduction}

% Optional reference for hybrid rendering: \cite{Beck:1981:HGR}
With recent advancements in GPU hardware acceleration, ray tracing is now feasible in real-time applications and is no longer limited to offline rendering. Hybrid rendering techniques that combine ray tracing and rasterization can generate higher quality visuals while maintaining interactive frame rates on PC applications. In contrast, the graphics capabilities of mobile and VR devices are weaker so the use of ray tracing can lead to undesirably low frame rates. Hence, current cloud gaming providers apply remote rendering by performing rendering on powerful cloud infrastructure and streaming the output to the users' access devices \cite{Holthe:2009:GLR}.

However, thin-client devices, albeit with lower computational power, can perform rasterization in real-time. Rather than solely relying on the server for rendering, we leverage this limited graphics capability of the client \cite{Cuervo:2015:KHM} via distributed rendering. In particular, our approach can help to alleviate the server's workload so it can focus on ray tracing. The ray-traced information can then be used to achieve better visual quality in the final output. 

Nonetheless, in order to combine the graphics capabilities of both the client and server to meet real-time performance constraints, we require excellent network bandwidth and latency for efficient two-way communication and data transfer. Fortunately, this can be achieved with the newest developments in 5G technology. With faster networks, we can improve the overall performance of ray tracing-incorporated real-time rendering, bringing more realistic graphics to thin-client games and VR applications.
\section{Design}

For distributed rendering, we adopt the UDP network protocol for efficient communication between the cloud server and client device. Nonetheless, UDP does not handle retransmission of lost packets. Retransmitted packets are also not useful for real-time interactive applications as they will be outdated for the current frame. Hence, in the event of a timeout, we make use of the latest available previously received data instead. 

As for hybrid rendering, we adapt the DirectX implementation of a simple hybrid rendering pipeline \cite{Wyman:2018:IDR} with Lambertian shading \cite{Koppal:LR:2014} and ray-traced shadows. We first perform a G-buffer rasterization pass for deferred shading to obtain the necessary per-pixel information required. Next, we perform ray tracing to query the relative visibility of every light in the scene with respect to each pixel. Lastly, we combine the G-buffer and light visibility information in computing the final pixel colour $I$ as shown.

\begin{equation}
  I = \frac{I_{d}}{\pi} \sum_{i} k_{i} \cdot \saturate{$N \cdot L_{i}$} I_{i}
\end{equation}

In the above formula, $I_{d} / \pi$ refers to the diffuse BRDF of the pixel with $I_{d}$ as its material diffuse colour. $N$ represents the pixel's surface normal while $k_{i}$, $L_{i}$ and $I_{i}$ refer to the relative visibility, direction and intensity of light $i$ respectively.

\subsection{Rasterization}

For each frame, the client first sends the dynamic updates in scene information including any user input and camera movement to the server for synchronization. Rasterization is then performed to obtain per-pixel information needed for the rest of the rendering. The server requires the pixels' world positions for the tracing of shadow rays, while the client requires their world space normal vectors and material diffuse colours for Lambertian shading.

For this particular pipeline, we perform rasterization on both the client and server at the same time. Alternatively, to avoid repeated work, we can also make the client or server perform rasterization and send the necessary information required by the other end over the network. If we let the client do the rasterization, we can leverage its limited computational capabilities and also minimize the amount of data that needs to be transferred (i.e. world positions only). On the other hand, rasterization on the server will most likely be faster because of its superior graphics hardware. The server can then proceed with ray tracing immediately rather than wait for the thin client to complete rasterization and send its data over.

Nonetheless, we choose to perform rasterization on both ends to improve the overall rendering performance. By performing rasterization on the server, we minimize the time taken to obtain the visibility bitmap. Although the client takes longer for rasterization, it is possible that the client would have completed rasterization by the time it receives the visibility bitmap from the server. It can then promptly proceed with Lambertian shading for the final output.

In this case, both the client and server require raster information. However, for other hybrid rendering pipelines, it would be better to avoid repeated work. For instance, \cite{Cabeleira:2010:CRR} introduces a pipeline that performs direct lighting computation in parallel with the ray tracing process. Such independent work distribution can enable our approach to attain better performance than pure remote rendering.

\subsection{Ray Tracing}

Shadow mapping is usually adopted to achieve real-time shadows in interactive applications. However, ray-traced shadows produce more accurate shadow boundaries by avoiding texture artifacts such as jaggies and shadow acne. Hence, we perform hardware-accelerated ray tracing on the server to incorporate high-quality ray-traced shadows in real-time.

For every pixel, we obtain its corresponding world position efficiently through rasterization as compared to ray casting from the pixel centre. From the world position, we then trace a shadow ray to every light in the scene. If this light-direction ray for light $i$ is not obstructed by any object, the light is visible from the pixel and $k_{i} = 1$. The pixel is hence deemed to be illuminated by the light. On the other hand, if the ray is obstructed by some object, $k_{i} = 0$ which indicates that the light is not visible from the pixel. Hence, the light does not contribute to the pixel's final colour.

This visibility boolean is stored in a frame-sized bitmap with a compact bitmask per pixel, where every bit represents the relative visibility of a light from the pixel. For every visible light, its direction $L_{i}$ and intensity $I_{i}$ are then included in the computation of the final pixel colour. The size of the per-pixel bitmask and number of overall visibility bitmaps can be increased based on the number of lights in the scene. To minimize the duration of visibility bitmap transfer, we apply LZ4 lossless compression to reduce the amount of data sent over the network. This duration can be further brought down with the help of high speed 5G networks.
\section{Discussion}

We tested our distributed hybrid rendering approach on \href{https://www.blendswap.com/blend/13491}{\textsc{The Modern Living Room}} (\href{https://creativecommons.org/licenses/by/3.0/}{CC BY}) with three lights. Hardware-accelerated ray tracing was performed on a PC with a GeForce RTX 2080 GPU. A reasonably interactive frame rate of 34 fps was achieved with our basic prototype without substantial optimization.

With more premium GPUs out in the market and the rapid development of 5G networks as well as edge computing, we believe that even higher frame rates can be attained with our current implementation. Nonetheless, we are continuing to explore optimizations in data transfer over the network as well as between local GPU and CPUs, and aim to enhance performance further with better data compression and transmission techniques. In the event of lost packets, we are also working on more accurate alternatives to the raw data of previous frames. For instance, we believe that we can use scene information like motion vectors in conjunction with history frames to estimate the contents of lost packets.

Eventually, we hope to bring our distributed rendering approach to more advanced pipelines and incorporate lighting and camera effects such as reflections, global illumination and depth of field.

\section*{Acknowledgements}

This work is supported by the Singapore Ministry of Education Academic Research grant T1 251RES1812, “Dynamic Hybrid Real-time Rendering with Hardware Accelerated Ray-tracing and Rasterization for Interactive Applications”. We thank developers Nicholas Nge and Alden Tan in our lab for their assistance.

%-------------------------------------------------------------------------
% bibtex
\bibliographystyle{eg-alpha-doi}
\bibliography{hybrid}

\newcommand{\etalchar}[1]{$^{#1}$}
\begin{thebibliography}{\uppercase{CWC{\etalchar{*}}15}}

\bibitem[Cab10]{Cabeleira:2010:CRR}
\textsc{Cabeleira J. P.~G.}:
\newblock \emph{Combining Rasterization and Ray Tracing Techniques to
  Approximate Global Illumination in Real-Time}.
\newblock Master's thesis, Portugal, Nov. 2010.
\newblock URL: \url{http://voltaico.net/files/article.pdf}.

\bibitem[CWC{\etalchar{*}}15]{Cuervo:2015:KHM}
\textsc{Cuervo E., Wolman A., Cox L.~P., Lebeck K., Razeen A., Saroiu S.,
  Musuvathi M.}:
\newblock Kahawai: High-quality mobile gaming using gpu offload.
\newblock In \emph{Proceedings of the 13th Annual International Conference on
  Mobile Systems, Applications, and Services} (New York, NY, USA, 2015),
  MobiSys '15, Association for Computing Machinery, p.~121–135.
\newblock URL: \url{https://doi.org/10.1145/2742647.2742657}, \href
  {https://doi.org/10.1145/2742647.2742657}
  {\path{doi:10.1145/2742647.2742657}}.

\bibitem[HMR09]{Holthe:2009:GLR}
\textsc{{Holthe} O., {Mogstad} O., {Ronningen} L.~A.}:
\newblock {Geelix LiveGames}: Remote playing of video games.
\newblock In \emph{2009 6th IEEE Consumer Communications and Networking
  Conference} (Jan 2009), pp.~1--2.
\newblock URL: \url{https://doi.org/10.1109/CCNC.2009.4784713}, \href
  {https://doi.org/10.1109/CCNC.2009.4784713}
  {\path{doi:10.1109/CCNC.2009.4784713}}.

\bibitem[Kop14]{Koppal:LR:2014}
\textsc{Koppal S.~J.}:
\newblock \emph{Lambertian Reflectance}.
\newblock Springer US, Boston, MA, 2014, pp.~441--443.
\newblock URL: \url{https://doi.org/10.1007/978-0-387-31439-6_534}, \href
  {https://doi.org/10.1007/978-0-387-31439-6_534}
  {\path{doi:10.1007/978-0-387-31439-6_534}}.

\bibitem[Wym18]{Wyman:2018:IDR}
\textsc{Wyman C.}:
\newblock Introduction to directx raytracing.
\newblock In \emph{ACM SIGGRAPH 2018 Courses} (Aug. 2018), SIGGRAPH ’18.
\newblock URL: \url{http://cwyman.org/code/dxrTutors/dxr_tutors.md.html}.

\end{thebibliography}

\end{document}